\documentclass{desyproc}

\input{paperdef}
\usepackage{amsmath,amssymb,color,cite}
\graphicspath{{figs/}}
\newcommand{\gmt}{$(g-2)_\mu$}
\newcommand{\bsg}{BR($b \to s \gamma$)}
\newcommand{\btn}{BR($B_u \to \tau \nu_\tau$)}
\newcommand{\bmm}{BR($B_s \to \mu^+\mu^-$)}

\newcommand{\Och}{\ensuremath{\Omega_\chi h^2}}
\newcommand{\ecm}{\sqrt{s}}

\begin{document}

\thispagestyle{empty}
\setcounter{page}{0}
\def\thefootnote{\fnsymbol{footnote}}

\begin{flushright}
\mbox{}
\end{flushright}

\vspace{1cm}

\begin{center}

{\large\sc {\bf Discovering SUSY in the First LHC Run}}
\footnote{talk given at the {\em Physics at the LHC 2010}, 
May 2010, DESY Hamburg, Germany}

\vspace{1cm}

{\sc 
S.~Heinemeyer%
\footnote{
email: Sven.Heinemeyer@cern.ch}%
}

\vspace*{1cm}

{\it
Instituto de F\'isica de Cantabria (CSIC-UC), 
Santander,  Spain\\

}
\end{center}

\vspace*{0.2cm}

\BC {\bf Abstract} \EC
We analyze the potential of the first LHC physics run, assuming 1~\ifb\
at $\ecm = 7 \tev$, to discover Supersymmetry (SUSY). 
The results are based on SUSY parameter fits following a frequentist
approach. They include the experimental constraints
from electroweak precision data, \gmt, $B$ physics and cosmological data.
The two SUSY models under consideration are the constrained MSSM (CMSSM) 
with universal soft supersymmetry-breaking mass parameters, and a model
with common non-universal Higgs mass parameters in the superpotential (NUHM1).
We find that large parts of the regions preferred at the 68\% C.L.\ are
accessible to early LHC running.

\def\thefootnote{\arabic{footnote}}
\setcounter{footnote}{0}

\newpage


\title{Discovering SUSY in the First LHC Run}


%
\author{{\slshape Sven Heinemeyer}\\[1ex]
Instituto de F\'isica de Cantabria (CSIC-UC), Santander, Spain}

%

%

\contribID{xy}  
\confID{1964}
\desyproc{DESY-PROC-2010-01}
\acronym{PLHC2010}
\doi            

\maketitle

\begin{abstract}
We analyze the potential of the first LHC physics run, assuming 1~\ifb\
at $\ecm = 7 \tev$, to discover Supersymmetry (SUSY). 
The results are based on SUSY parameter fits following a frequentist
approach. They include the experimental constraints
from electroweak precision data, \gmt, $B$ physics and cosmological data.
The two SUSY models under consideration are the constrained MSSM (CMSSM) 
with universal soft supersymmetry-breaking mass parameters, and a model
with common non-universal Higgs mass parameters in the superpotential (NUHM1).
We find that large parts of the regions preferred at the 68\% C.L.\ are
accessible to early LHC running.

\end{abstract}


\section{Introduction}

One of the main tasks of the LHC is to search for physics beyond the
Standard Model (SM), where Supersymmetry (SUSY) is one of the favored
ideas. The first physics run of the LHC is currently ongoing at an 
$\ecm = 7 \tev$, aiming for 1~\ifb\ until the end of 2011.
Here we review the results from frequentist
analyses~\cite{Master3,Master2} of 
the parameter spaces of the constrained minimal supersymmetric
extension of the Standard Model (CMSSM) --- in which the
soft SUSY-breaking scalar and gaugino masses are each constrained
to universal values $m_0$ and $m_{1/2}$, respectively 
 --- and the NUHM1 --- in which the soft SUSY-breaking
contributions to the Higgs masses are allowed to have a different but
common value. Both models have a common trilinear coupling $A_0$ at the
GUT scale and $\tb$ (the ratio of the two vacuum expectation values) as
a low-energy input. 
A detailed list of references on the subject of frequentist (and
bayesian) analyses can be found in \citere{Master3}.


\section{Details of the fits}

The results obtained in \citeres{Master3,Master2} include various
experimental results: 
$B$-physics observables (such as rates for \bsg\ and \btn,
and the upper limit on \bmm) as well as $K$-physics observables,
precision electroweak data (such as the $W$~boson mass and the anomalous
magnetic moment of the muon, \gmt), 
the bound on the lightest MSSM Higgs boson mass, $\Mh$, 
and the cold dark matter (CDM) density
(as inferred from astrophysical and cosmological data) 
assuming that this is
dominated by the relic density of the lightest neutralino, $\Och$. 
 
The fit is performed by using 
a global $\chi^2$ likelihood function, which combines all
theoretical predictions with experimental constraints:
\begin{align}
\chi^2 &= \sum^N_i \frac{(C_i - P_i)^2}{\sigma(C_i)^2 + \sigma(P_i)^2}
\nonumber 
+ \sum^M_i \frac{(f^{\rm obs}_{{\rm SM}_i}
              - f^{\rm fit}_{{\rm SM}_i})^2}{\sigma(f_{{\rm SM}_i})^2}
                                                  \non \\[.2em]
&+ {\chi^2(\Mh) + \chi^2(\br(B_s \to \mu\mu))}
+ {\chi^2(\mbox{SUSY search limits})}
\label{eqn:chi2}
\end{align} 
Here $N$ is the number of observables studied, $C_i$ represents an
experimentally measured value (constraint), and each $P_i$ defines a
prediction for the corresponding constraint that depends on the
supersymmetric parameters.
The experimental uncertainty, $\sigma(C_i)$, of each measurement is
taken to be both statistically and systematically independent of the
corresponding theoretical uncertainty, $\sigma(P_i)$, in its
prediction. We denote by
$\chi^2(\Mh)$ and $\chi^2(\br(B_s \to \mu\mu))$ the $\chi^2$
contributions from two measurements for which only one-sided
bounds are available so far. Similarly, 
we include the lower limits from the direct searches
for SUSY particles at LEP~\cite{LEPSUSY} as one-sided limits, denoted by 
``$\chi^2(\mbox{SUSY search limits})$'' in \refeq{eqn:chi2}.
The experimental constraints used in our analyses are listed in
Table~1 of \citere{Master3}. 
Our statistical treatment of the CMSSM and NUHM1 makes use of  
a large sample of points (about $2.5 \times 10^7$) 
in the SUSY parameter spaces 
obtained with the Markov Chain Monte Carlo
(MCMC) technique.
Our analysis is entirely frequentist, and avoids any
ambiguity associated with the choices of Bayesian priors.


The main computer code for our evaluations is the 
{\tt MasterCode}~\cite{Master1,Master2,Master3,Master35,MasterWWW}, 
which includes the following theoretical codes. For the RGE running of
the soft SUSY-breaking parameters, it uses
{\tt SoftSUSY}~\cite{Allanach:2001kg}, which is combined consistently
with the codes used for the various low-energy observables:
{\tt FeynHiggs}~\cite{Heinemeyer:1998yj,Heinemeyer:1998np,Degrassi:2002fi,Frank:2006yh}  
is used for the evaluation of the Higgs masses and  
$a_\mu^{\rm SUSY}$  (see also
\cite{Moroi:1995yh,PomssmRep}),
for the other electroweak precision data we have included 
a code based on~\cite{MWMSSM,ZObsMSSM},
{\tt SuFla}~\cite{Isidori:2006pk,Isidori:2007jw} and 
{\tt SuperIso}~\cite{Mahmoudi:2008tp,Eriksson:2008cx}
are used for flavor-related observables, 
and for dark-matter-related observables
{\tt MicrOMEGAs}~\cite{Belanger:2006is} and
{\tt DarkSUSY}~\cite{Gondolo:2005we} are included.
In the combination of the various codes,
{\tt MasterCode} makes extensive use of the SUSY
Les Houches Accord~\cite{Skands:2003cj,Allanach:2008qq}.


\section{SUSY discovery potential of the first LHC run}

For the parameters of the best-fit CMSSM point we find
$m_0 = 60 \gev$,  $m_{1/2} = 310 \gev$,  $A_0 = 130 \gev$, $\tb = 11$
and $\mu = 400 \gev$, 
yielding the overall $\chi^2/{\rm N_{\rm dof}} = 20.6/19$ (36\% probability) 
and nominally $\Mh = 114.2 \gev$.
The corresponding parameters of
the best-fit NUHM1 point are $m_0 = 150 \gev$, $m_{1/2} = 270 \gev$,
$A_0 = -1300 \gev$, $\tb = 11$ and
$m_{h_1}^2  = m_{h_2}^2 = - 1.2 \times 10^6 \gev^2$ or, equivalently,
$\mu = 1140 \gev$, yielding
$\chi^2 = 18.4$ (corresponding to a similar fit probability as in the CMSSM)
and $\Mh = 120.7 \gev$. 

In Fig.~\ref{fig:contours} we display the best-fit value and the 
68\% and 95\% likelihood contours 
for the CMSSM (upper plot) and the NUHM1 (lower plot) in the 
$(m_0, m_{1/2})$ plane, obtained as described above from a fit taking into
account all experimental constraints.
We also show exclusion contours for the hadronic search mode (jets plus
missing energy) at CMS. 
The green (light gray) solid line shows the 95\%~C.L.\ exclusion contour
for CMS for 1~\ifb\ at $\ecm = 7 \tev$~\cite{CMS7tev}. The black solid
line shows the corresponding results for only 0.1~\ifb.
(Similar results hold for ATLAS.)
One can see that with 1~\ifb\ the best-fit points can be tested,
together with a sizable part of the whole 68\%~C.L.\ preferred
regions. In the case of the NUHM1 (lower plot) nearly the 68\%~C.L.\
region could be probed.

\begin{figure*}[htb!]
\begin{center}
\includegraphics[width=0.75\textwidth]{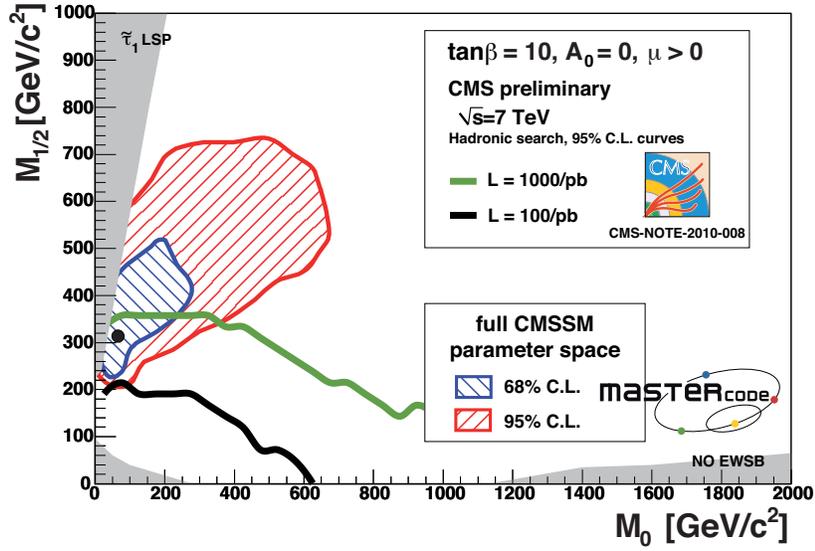}\\[1em]
\includegraphics[width=0.75\textwidth]{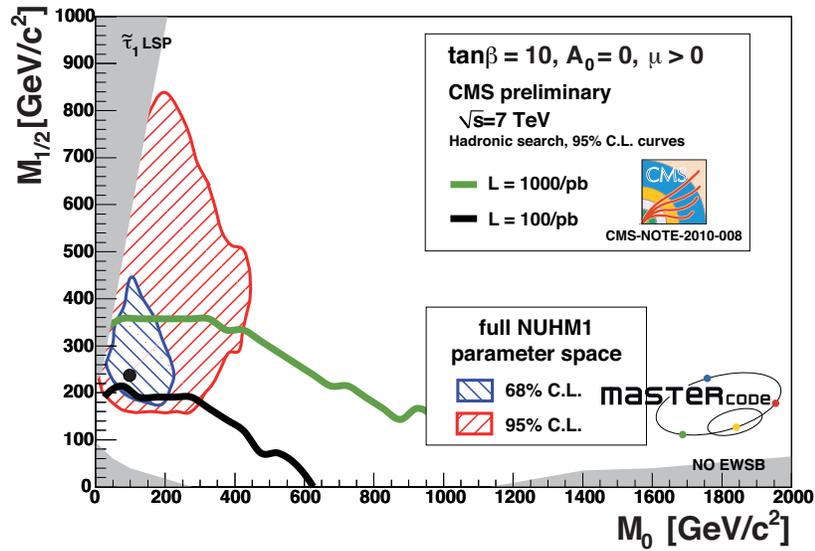}
\end{center}
\vspace{-1.5em}
\caption {The $(m_0, m_{1/2})$ plane in the CMSSM (upper plot) and the
  NUHM1 (lower plot).
  The dark shaded area at low $m_0$ and high $m_{1/2}$ is
  excluded due 
  to a scalar tau LSP, the light shaded areas at low $m_{1/2}$ do not
  exhibit electroweak symmetry breaking. 
  Shown in both plots are the best-fit point, indicated by a filled
  circle, and the 
  68 (95)\%~C.L.\ contours from our fit as dark gray/blue (light
  gray/red) overlays, scanned over all $\tan\beta$ and $A_0$ values.
  The 95\% C.L.\ exclusion curves (hadronic search channel) 
  at CMS with 1~(0.1)~fb$^{-1}$ at 7~TeV center-of-mass energy 
  is shown as green/light gray (black) solid curve.
} 
\label{fig:contours}
\end{figure*}

In conclusion, if the CMSSM or the NUHM1 (or a very similar SUSY model)
were realized in nature, the first LHC physics run at $\ecm = 7 \tev$
until the end of 2011 could reveal already first signals of SUSY. On the
other hand, no indication of SUSY-like signatures would already severly
restrict these (kind of) GUT based models.


\section{Acknowledgments}

We thank O.~Buchm\"uller, R.~Cavanaugh, A.~De~Roeck, J.~Ellis,
H.~Fl\"acher, G.~Isidori, K.~Olive, F.~Ronga and G.~Weiglein 
with whom the results presented here have been obtained.
Work supported in part by the European Community's Marie-Curie Research
Training Network under contract MRTN-CT-2006-035505
`Tools and Precision Calculations for Physics Discoveries at Colliders'
(HEPTOOLS).




\begin{footnotesize}

\newcommand{\ea}{{\it et al.}}

\end{footnotesize}


\end{document}